\documentclass [12pt,english]{article}
\pdfoutput=1
\usepackage{amssymb}
\usepackage{amsmath}
\usepackage{amsfonts}
\usepackage[english]{babel}

\usepackage[T2A]{fontenc}
\usepackage[cp1251]{inputenc}
\usepackage{fullpage}
\usepackage{bm}
\usepackage{graphicx}
\numberwithin{equation}{section}
\newcounter{Alpha}

\begin{document}

\title{\LARGE 
%\Roman{0}
D--dimensional half--filled Hubbard model.\\
Zero temperature  paramagnetic solution}

\author{N. I. Chashchin\thanks{E--mail: nik.iv.chaschin@mail.ru} \\ \textit{Ural State Forestry University}\\
%, Russian Academy of Sciences}
\textit {Ekaterinburg, Sibirskii trakt 37, 620100 Russia}}
\date{}

\maketitle

\begin{abstract}
In this paper we investigate effects of a lattice dimension on strongly correlated 
electronic systems at $T=0 K$. The model for numerical calculations is formalized 
in terms of the integral equations which were obtained previously for the half -- filled 
paramagnetic Hubbard model.  The assumption that all relevant functions of 
the system have a magnitude dependence of momentus vectors makes ultimately possible treating 
a lattice dimension $D$ as an external parameter.  We here submit for consideration 
the number of doubly occupied lattice sites, the electronic density of states, 
and the momentum distribution in electronic occupation numbers at $D$=1, $D$=2, $D$=3. 
\end{abstract}

\\
{}
\\
DOI:10.1134/S0031918X1607036\\
Pacs numbers 31.15.xt; 31.15.xp; 71.10.Fd

\section{\!\!\!\!\!\!. Introduction}

The Hubbard model (HM) \cite{Hubbard63} long ago had been acknowledged as the archetypal model for theoretical studying of strongly correlated systems. 
%and their phase transitions in condensed matter physics.
Originally, the Hubbard model was introduced %\cite{Hubbard1963} 
for describing electronic correlations in narrow energy bands of transition metals but subsequently it turned out   
to be a very simple pattern and benchmark for understanding of the unusual electronic properties of the strongly correlated systems. A great number of remarkable dynamical phenomena have been reported, including a correlation---driven Mott transition, band collapsing, and mass enhancement (heavy fermions) at the Fermi level; as well as non-Fermi-liquid properties, a pseudogap in the electronic spectrum, and so on. It is indicative that angle--resolved photoemission studies (ARPES) of materials report unusual quasiparticle properties at all carrier concentrations. The apparent strong coupling nature of the problems suggests a nonperturbative treatment is required. 

The one--dimensional (1D) HM is much easier for the considiration. In 1968  Lieb and Wu published the exact solution of the problem based on the the Bethe ansatz \cite{Lieb_Wu}. 
They showed that half -- filled HM has an antiferromagnetic ground state without the Mott transition for any values of the one--site Coulomb electron--electron interaction. Impressive results were achieved by so called bosonization method \cite{{Benthien},{Sing},{vonDelft}}. The idea of the bosonization is based on the
construction of a complete set of some bosonic operators  which then diagonalize 1D fermionic Hamiltonian. 

Further progress in the nonperturbative describing of HM was connected with the development of the 
dynamic meanfield method (DMFT) and its extensions \cite{{Metzner89},{Georges96}}. The method is rely upon the observation that in the limit of the infinite--dimensional lattice the electronic self--energy proves to be a local function. 
In this case, the HM is equivalent to the single--impurity Anderson model which is used as a solver. 
Thus, DMFT method becomes exact at D$\,\to\!\infty$ and yields a close approximation for the finite dimensional lattice. It has been recognized DMFT a big step forward in the understanding of strongly correlated electronic systems.

One might say basic properties of the 1D and the high--dimensional HM for the most part are known, 
but in intermediate dimensions the problems such as 
a correlation--driven Mott-Hubbard insulator transition (MIT), a pseudogap in the electronic spectrum, 
charge and spin density waves caused by the magnetic and charge ordering, existence of the non--Fermi--liquid  
behavior of the electronic system at D>1 still remain under question.  

Ordinary, multi--dimensional metals are described by Fermi liquid theory.
%Fermi liquid theory is a key point of electron--electron interactions in metals. 
It states there is one--to--one correspondence between the 
low--energy excitations of a free Fermi gas and quisiparticles of an interacting electronic liquid. 
The MIT is a demonstrative example of the strongly  correlated many--electronic 
phenomenon, where the quisiparticle  assumption fails at least at low dimensions.

2D and 3D HM has long been a playground for numerical simulations 
\cite{{Tudor},{2D_Igoshev},{3D_Khatami}}, not to mention of various  perturbative methods, and it is clear that lattice dimension appears to be a significant parameter for system's behavior and a further considiration is needed.  

In our previous works \cite{{chaschin2011_2},{chaschin2012_3},{chaschin2016_4},{chaschin2011_1},chaschin2017} we had obtained    
by means of generating functional  method and the subsequent Legendre transformation an extended  mathematical model of HM (see Appendix). After some transformation the dimension $D$ shows up in the model integral equations as an external parameter. Hence, we get resources to study directly   
an influence of the lattice dimension on some correlation phenomena; we here submit for consideration the number of double--occupied sites, the electronic density of states, and the momentum distribution functions. 
 
%The rest of the paper is organized as follows. In section I  
%full panoply -- ��� ������������

\section{\!\!\!\!\!\!. Model and Method} 

%In this article our intention is to consider the half--filled D--dimensional HM. 

In the simplest form, the Hamiltonian of the Hubbard model is written as

\begin{equation}
\mathcal{H}= -t\sum\limits_{\langle i,j\rangle\sigma}c_{i\sigma}^{\dag}c_{j\sigma}+\sum\limits_{i\sigma}
\varepsilon_\sigma n_{i\sigma}
+U\sum\limits_{i}n_{i\uparrow}n_{j\downarrow}\,,
\label{Pm:Hub_ham}
\end{equation} 
where $U$ is the parameter of the Coulomb interaction at a site;  
$c_{i\sigma} (c_{i\sigma}^\dag)$ are the Fermi operators describe 
the annihilation (generation) of electrons with spins up and down $\sigma=\uparrow,\downarrow$; 
 $n_{i\sigma}$  indicates the operators of the number of particles; $t$ is the parameter 
of hopping of s--electrons from site to site; in the designation
$\langle i,j\rangle$ sites are nearest; $\varepsilon_\sigma=-\displaystyle{\sigma\frac{h}{2}-\mu}$,
where $h=g\mu_BH$ and $g$ is the electronic $g-$factor, $\mu_B$ is the Bohr magneton; 
$H$ is the external magnetic field, and $\mu$ is the chemical potential.

As it is shown in Appendix 
we had got two type of propagators: the fermionic (A.6) one ($N=G_\uparrow+G_\downarrow=2\,G$ in the paramagnetic case) 
   
\begin{equation}
\displaystyle N({\bf k}, i\omega_n)=\frac{2}{i\omega_n-\varepsilon_k-\Sigma({\bf k},i\omega_n)}\,,
\label{N_el}
\end{equation}
where $\omega_n=(2 n+1)\pi T$ ($n=0,\pm 1,\pm 2,\dots$) is electronic  Matsubara frequencies, $\Sigma({\bf k},i\omega_n)$ is the self energy, and $\varepsilon_k$ is the spectrum of free electrons;\\ 
and the bosonic charge propagator (A.5)  
\begin{equation}
\displaystyle Q({\bf q},i\Omega_\nu)=-\frac{1}{1+\frac{U}{2}\Pi(q,i\Omega_\nu)}\,,
\label{Q_bos}
\end{equation}
where $\Omega_\nu=2\, \nu \,\pi T$ ($\nu=0,\pm 1,\pm 2,\dots$) is bosonic Matsubara frequencies and 
$\Pi({\bf q},i\Omega_\nu)$ --- the bosonic self energy.  

For the D--dimensional cubic lattice the momentum vector $\bf k$, the momentum magnitude $k$, and the free electronic spectrum $\varepsilon({\bf k})$ in the strong--coupling regime are 
 
\begin{equation}
\begin{array}{l}
\displaystyle {\bf k}=(k_1,k_2,\dots k_D),\,\, k=\sqrt{k^2_1+k^2_2+\dots+k^2_D}\:;
\\ 
\displaystyle \varepsilon({\bf k})=-2\,t\sum\limits_{i=1}^{D}\cos(k_i)\,.
\end{array}
\label{Spectrum_D}   
\end{equation}

We choose the free electronic energy spectrum $\varepsilon_k$ as a function of the momentum magnitude $k$ 
(\ref{Spectrum_D}) (supposing henceforth 2t = 1)

\begin{equation}
\displaystyle \varepsilon_k=-D\cos\left(\frac{\pi}{K_D}k\right)\,,
\label{D_freespect}
\end{equation}
where parameter $K_D$ is radius of the first Brillouin zone (BZ) will be defined below in (\ref{PM:Kd}). 
The structure of Eqs. (\ref{Ap:PI_Q}, \ref{Ap:Sigma_N}, \ref{Ap:Symm}) allows to take an obvious 
assumption that in this case all relevant functions of our model become ${\bf k}^2-$dependent through $\varepsilon(k)$, and that enables to represent lattice sums as the following well known in any advanced course of Mathematical Analysis $D-$space sphere integrals: 
%\cite{D_Sphere}:
\begin{equation}
\begin{array}{l}
\displaystyle \sum\limits_{{\bf k}}f({\bf k}^2) = \int_{-\pi}^\pi \frac{d^D k}{(2\pi)^D}\,f({\bf k}^2)=  
\displaystyle C_D\int_{0}^{K_D} k^{D-1}f(k^2)\,d k\,,\\{}\\
\displaystyle \sum\limits_{{\bf k}}f({\bf k}^2,{\bf k} {\bf q}) =  
\displaystyle \frac{C_{D-1}}{2\pi}\int_{0}^{K_D} k^{D-1}\int_{0}^{\pi}\,f(k^2,k q \cos \theta)
(\sin\theta)^{D-2}d k\,d \theta\,,
\end{array}
\label{D_k}
\end{equation}
(for the matter of correctness these formulas are rigorous only for $K_D\gg 1$)
where factors $C_D$ are determined by the gamma function $\Gamma$ as 
\begin{equation}
C_D^{-1}=2^{D-1}\,\pi^{D/2}\,\Gamma(D/2)\:;\quad
C_1=\frac{1}{\pi},\, C_2=\frac{1}{2\pi},\,C_3=\frac{1}{2\pi^2}\,,\dots\;.
\label{CD}
\end{equation}

The completness condition of the first BZ volume yields a value of its radius $K_D$ by the following 
consideration: 
\begin{equation}
\displaystyle \sum\limits_{{\bf k}}1 = \int_{-\pi}^\pi \frac{d^D k}{(2\pi)^D}\,=  
\displaystyle C_D\int_{0}^{K_D} k^{D-1}d k = 1\,; \quad K_D=(\frac{D}{C_D})^{1/D}\,,
\label{PM:Kd}
\end{equation}
i.e., particularly $K_1=\pi,\, K_2=2\,\pi^{1/2},\,K_3=(6\,\pi^2)^{1/3},\dots$\,. The $D-$dimension Fermi momentum is  
\begin{equation}
\displaystyle k_{FD}=\frac{1}{2}K_D\,. 
\label{PM:KF}
\end{equation}
For example, $k_{F1}\approx$1.57, $k_{F2}\approx$1.77, $k_{F3}\approx$1.95\,,\dots\;.
Taking into account (\ref{D_k}, \ref{PM:Kd}) and transforming Eqs.\,(\ref{Ap:Sigma_N}) under symmetry condition (\ref{Ap:Symm}), we get the electronic part of the model equations  

\begin{equation}
\left\{
\begin{array}{l}
\displaystyle \Im \Sigma(k,\omega) = \frac{U C_{D-1}}{4\pi}\int_{0}^{K_D}\!q^{D-1}
\displaystyle\left[1-\tanh(\frac{\varepsilon_q}{2 T})
\displaystyle \tanh(\frac{\varepsilon_q-\omega}{2 T})\right]
\\{}\\
\times
\displaystyle\int_0^\pi(\sin\theta)^{D-2}\,
\Im Q\left(\sqrt{q^2+k^2-2\,q\,k\cos\theta}\;,\varepsilon_q-\omega\right) d q\,d\theta\,,
\\
{}
\\
\displaystyle \Re \Sigma(k,\omega) = \frac{1}{\pi}\int_{-\infty}^\infty  
\displaystyle \frac{\Im \Sigma(k,\omega^\prime)}{\omega^\prime - \omega}\,
d\omega^\prime\,;
\\
{}
\\
\Im N(k,\omega)=\displaystyle\frac{2\,\Im\Sigma(k,\omega)}
{\left[\omega-\varepsilon_k-\Re\Sigma(k ,\omega)\right]^2+
\left[\Im \Sigma(k;\omega)\right]^2}\,,
\\
{}
\\
\displaystyle \Im\Sigma(k+K_D,-\omega) = \Im\Sigma(k,\omega), 
\quad \Re\Sigma(k+K_D,-\omega) = -\Re\Sigma(k,\omega)\,,
\\{}\\
\displaystyle \Im N(k+K_D,-\omega) = \Im N(k,\omega)\,.
\end{array}
\right.
\label{Sigma}
\end{equation}
Here we connect $\Re \Sigma(k,\omega)$ and $\Im \Sigma(k,\omega)$ by means of the spectral Kramers--Kr\"{o}nig relation. 
  
Yet note that all temperature dependences of the model is confined by a unique factor 
\[
\displaystyle\left[1-\tanh(\frac{\varepsilon_q}{2 T})
\displaystyle \tanh(\frac{\varepsilon_q-\omega}{2 T})\right]\,,
\]
what permits us to get solutions at different temperatures. 

In our case of the paramagnetic zero temperature solution the bosonic --- charge excitation part of the model is obtained similarly to (\ref{Sigma}), using Eqs.\,(\ref{Ap:PI_Q}) 

\begin{equation}
\left\{
\begin{array}{l}
\displaystyle \Im \Pi(q,\Omega) = \frac{U C_{D-1}}{4\pi}\int_{0}^{K_D}
\displaystyle k^{D-1}\left[1-\tanh(\frac{\varepsilon_k}{2 T})
\displaystyle \tanh(\frac{\varepsilon_k-\Omega}{2 T})\right]
\\{}\\
\times
\displaystyle\int_0^\pi(\sin\theta)^{D-2}\:
\Im N\left(\sqrt{k^2+q^2-2\,k\,q\cos\theta}\;,\varepsilon_k-\Omega\right) d k\,d\theta\,,
\\
{}
\\
\displaystyle \Re \Pi(q,\Omega) = \frac{1}{\pi}\int_{-\infty}^\infty  
\displaystyle \frac{\tanh(\frac{\Omega^\prime}{2 T})\Im \Pi(q,\Omega^\prime)}{\Omega^\prime - \Omega}\,
d\Omega^\prime\,;
\\
{}
\\
\Im Q(q,\Omega)=\displaystyle\frac
{2 \,\Im \Pi(q,\Omega)}{\left[2+\Re\Pi(q,\Omega)\right]^2+
\left[\Im \Pi(q,\Omega)\tanh(\frac{\Omega}{2 T})\right]^2}\,; 
\\
{}
\\
\displaystyle \Im\Pi(q+K_D,-\Omega) = \Im\Pi(q,\Omega), 
\quad \Re\Pi(q+K_D,-\Omega) = \Re\Pi(q,\Omega)\,,
\\{}\\
\displaystyle \Im Q(q+K_D,-\Omega) = \Im Q(q,\Omega)\,.
\end{array}
\right.
\label{Pi_Q}
\end{equation}
\\
The coupled equations (\ref{Sigma}, \ref{Pi_Q}) present a model for numerical calculation of the 
 $D>1$~--~dimensional paramagnetic HM, where now the parameter $U C_{D-1}$ plays a role of an interaction constant. 
The case  $D=1$ had considered separately in \cite{{chaschin2016_4}, {chaschin2017}}. 
The lattice dimension enters into the model as an external parameter, and it is not necessarily an integer number. From Fig.1 we see that $C_{D}\ll1$ for D>4 therefore, correlation effects for large dimensions are perceptible only if Coulomb interaction $U$ is quite large. 

\begin{figure}[ht]
\begin{center}
\includegraphics[width=0.4\textwidth]{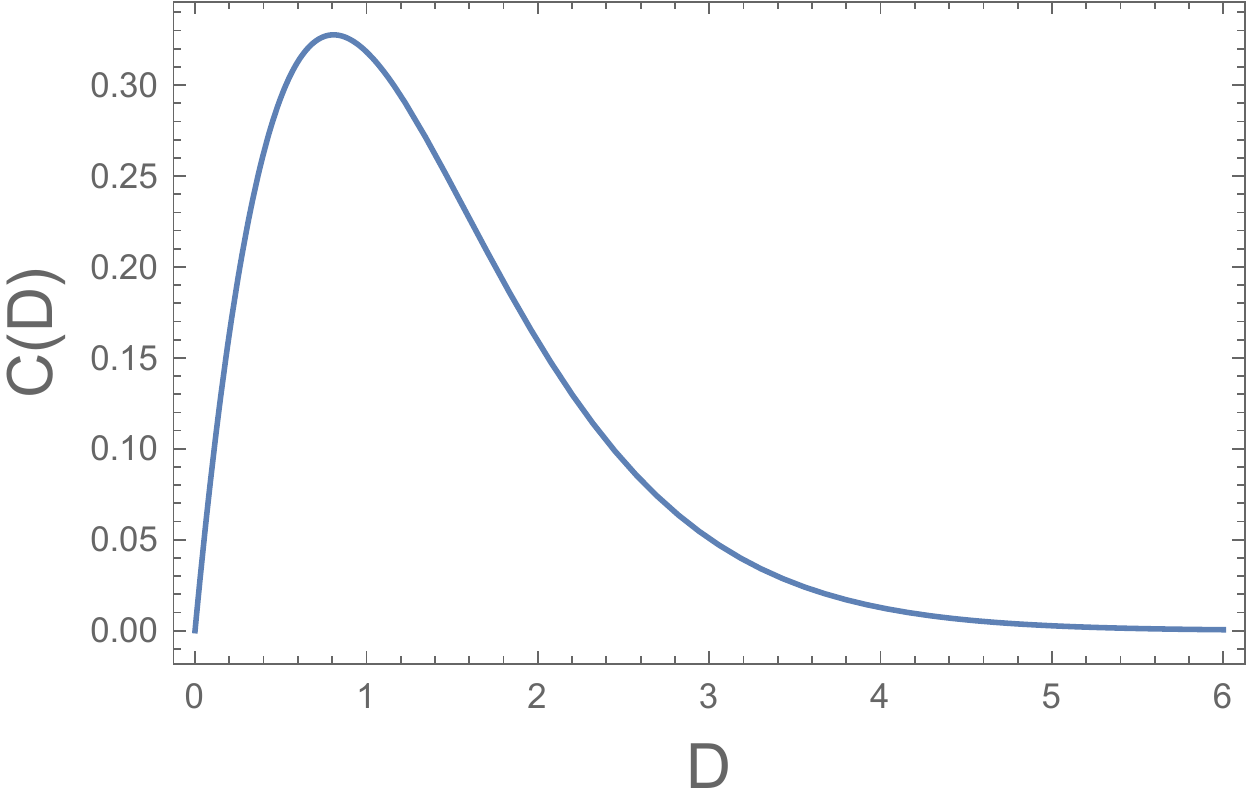}
\caption{C(D) defines an interaction constant as a function of the lattice dimension. 
We see that $C_{D}\approx 0$ for $D>4$ and hence, $D=4$ can be marked as some number threshold for correlation effects.} 
\label{Pm:Cd}
\end{center}
\end{figure} 

Thus, we get two coupled sets of the integral equations (\ref{Sigma}, \ref{Pi_Q}), which can be numerically calculated and investigated.   
Notice that they are structurally identical over the mutual substitutions: ${\Im N}\leftrightarrow{\Im Q}$, ${\Im\Sigma}\leftrightarrow{\Im\Pi}$, and ${\Re\Sigma}\leftrightarrow{\Re\Pi}$. 
%are proved for arbitrary $D$ and $T=0K$.    

\section{\!\!\!\!\!\!. Results and Discussion}

\addtocounter{Alpha}{1}
{\bf\large{\Alph{Alpha}. Double occupancy}}\\ %\setcounter{equation}{1}
\setcounter{Alpha}{1}
\addtocounter{Alpha}{1}

According to the formulas obtained in \cite{chaschin2011_1}, we have ($1\equiv i,\tau$)
\begin{equation}
\displaystyle \langle \hat T\, m_1\,m_2\rangle=\frac{\delta^2\Phi}{\delta \eta(11)\,{\delta\eta(22)}}\,,
\label{DPM:Xm}   
\end{equation}
where $\Phi=\ln Z$ is the generated functional of the connected Green functions; $\eta(12)=h\,\delta_{12}$ is the external magnetic field; $m_1=n_{1\uparrow}-n_{1\downarrow}$ --- 
the local magnetic moment, and $\hat T $ is the imaginary time ordering operator in the sym--form \cite{chaschin2011_1}: 
$\hat T n_1=n_1-1/2$.   
From this it follows that 
\begin{equation}
\displaystyle \langle\hat T\, m^2\rangle=\frac{1}{2} -2\,\langle n_{\uparrow}n_{\downarrow}\rangle\,. 
\label{DPM:X}   
\end{equation}

\begin{figure}[ht]
\begin{center}
\includegraphics[width=0.4\textwidth]{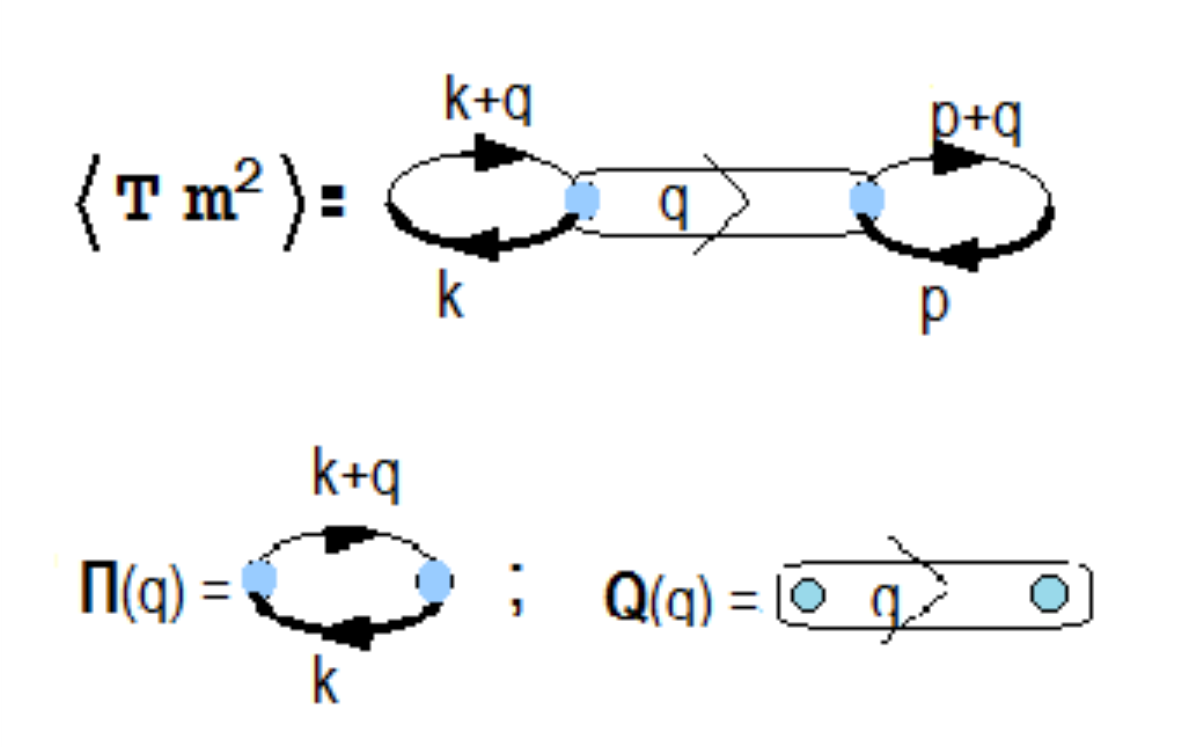}
\caption{Diagrammatic representation of the local $\langle T m^2\rangle$--correlator. 
The thin lines are free propagators $G_0$, the thick ones --- propagators of the correlated particles $N=2\,G$. The repeated indexes are supposed to be summed; here $k\equiv(k, i\omega_n)$.}
\label{Pm:XQ}
\end{center}
\end{figure} 

According to the diagrammatical representation (Fig.2) and the subsequent standart transformation, after taking into account the denotation 
$\Im\Pi({\bf q}, \Omega)=-\,\coth(\frac{\Omega}{2 T})\Im\Pi({\bf q}, \Omega+i 0)$,  
 we have 
\begin{equation}
\displaystyle \langle\hat T\, m^2\rangle = \frac{U}{2}\sum\limits_{{\bf q},\Omega_\nu}
\displaystyle\Pi^2_{\bf q}(i\Omega_\nu) Q_{\bf q}(i\Omega_\nu)= 
\displaystyle \frac{C_D}{\pi U}\int_{0}^{K_D} q^{D-1}\int_{-\infty}^\infty  
\left[\frac{1}{2}\Im\Pi_{q}(\Omega)-\Im Q_{q}(\Omega)\right] d q\, d \Omega\,,
\label{DPM:X_calc}
\end{equation}
where $\Im\Pi_{q}(\Omega)$ and $\Im Q_{q}(\Omega)$ are calculated in (\ref{Pi_Q}). 

Finally, we get from (\ref{DPM:X}) the average number of double--occupied sites  
\begin{equation}
\displaystyle \langle n_{\uparrow}n_{\downarrow}\rangle=\frac{1}{4}-\frac{\langle\hat T\, m^2\rangle}{2}\,.   
\label{DPM:Dbl}   
\end{equation}

Fig.3 represents graphics of the number of double--occupied sites $\langle n_{\uparrow}n_{\downarrow}\rangle$   ("twos") as a function of Coulomb interaction $U$ at the lattice dimensions: $D=1, D=2, D=3$.
Correlation--driven electronic charge density waves (CDW) are dissolving the local "twos" and their disordered lattice structure. Values of Coulomb interaction $U_{cD}$ correspond to 
$\langle n_{\uparrow}n_{\downarrow}\rangle$= 0 for each $D$. A negative value of the double occupancy parameter for $U>U_{cD}$ implies a CDW---instability that brings the lattice to a period doubling with respect to local charge values. As it was in the case of the magnetic dimerization at $D=1$ \cite{chaschin2016_4} this lattice distortion is accompanied by the appearance of a finite gap in the electronic spectrum, which means a phase transition from a semimetal with a pseudogap to an isolator with $\langle n_{\uparrow}n_{\downarrow}\rangle$ as an order parameter.\\
 
 \begin{figure}[ht]
\begin{center}
\includegraphics[width=0.45\textwidth]{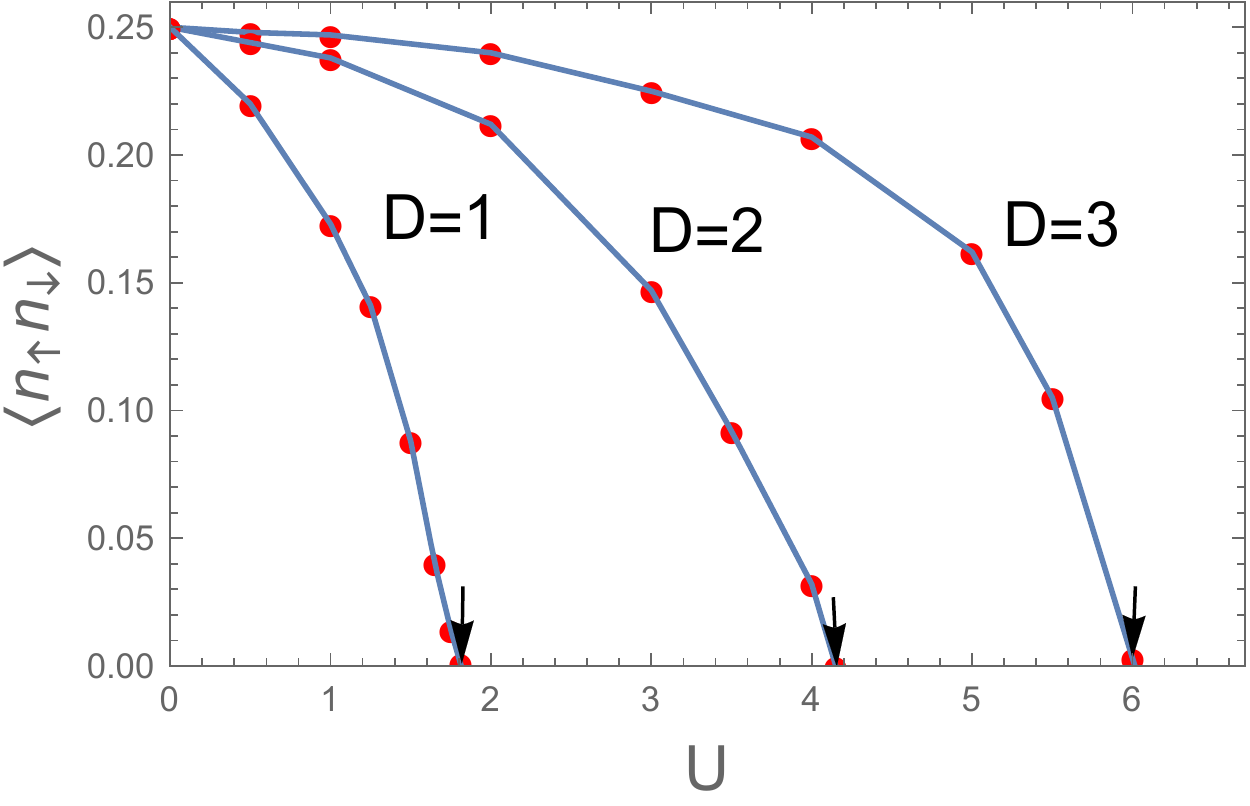}
\caption{The number of double--occupied sites versus $U$ 
at $D$=1, $D$=2, $D$=3. The arrows indicate numerical values of $U_{cD}$ whenever 
$\langle n_{\uparrow}n_{\downarrow}\rangle=0$; $U_{c1}\approx$ 1.8, $U_{c2}\approx$ 4.1, $U_{c3}\approx$ 6.1.}
\label{Pm:Dbl}
\end{center}
\end{figure} 
 
{\bf\large{\Alph{Alpha}. Density of one-electronic states}}\\ %\setcounter{equation}{1}
\setcounter{Alpha}{2}
\addtocounter{Alpha}{2} 

The density of electronic states (DOS) is determined in ordinary way  
\begin{equation}
\displaystyle \rho(\omega) =-\frac{1}{\pi}\sum\limits_{\bf k}\Im N(k;\omega)=
-\frac{C_D}{\pi}\int_{0}^{K_D} \,k^{D-1}\:\Im N(k;\omega)\,d k\,,
\label{DPM:rho_el}
\end{equation} 
where the imaginary part of the electronic Green function $\Im N=2\,\Im G$  is calculated in 
(\ref{Sigma}). 

The momentum distribution function $n(k)$ that gives an average occupation number of electronic states with momentum $k$ reads   
\begin{equation}
\displaystyle n(k)= -\frac{1}{2\pi}
\displaystyle\int_{-\infty}^\infty f_F\left(\frac{\omega}{2 T}\right)\,\Im N(k,\omega)\, d \omega\,,  
\label{DPM:nk}
\end{equation}
where $f_F$ is the Fermi function.

The relevance of function (\ref{DPM:nk}) refers to its analitical properties. For example, a presence or absence of a finite jump at the Fermi momentum $k_F$ are generally considered for discerning among different sorts of an electronic behavior; the finite jump would indicate that the qusiparticle excitations are of the Fermi liquid type without any gap in the spectrum and therefore, the system is of a metallic type.

\begin{figure}[ht]
\begin{center}
 mode\includegraphics[width=0.8\textwidth]{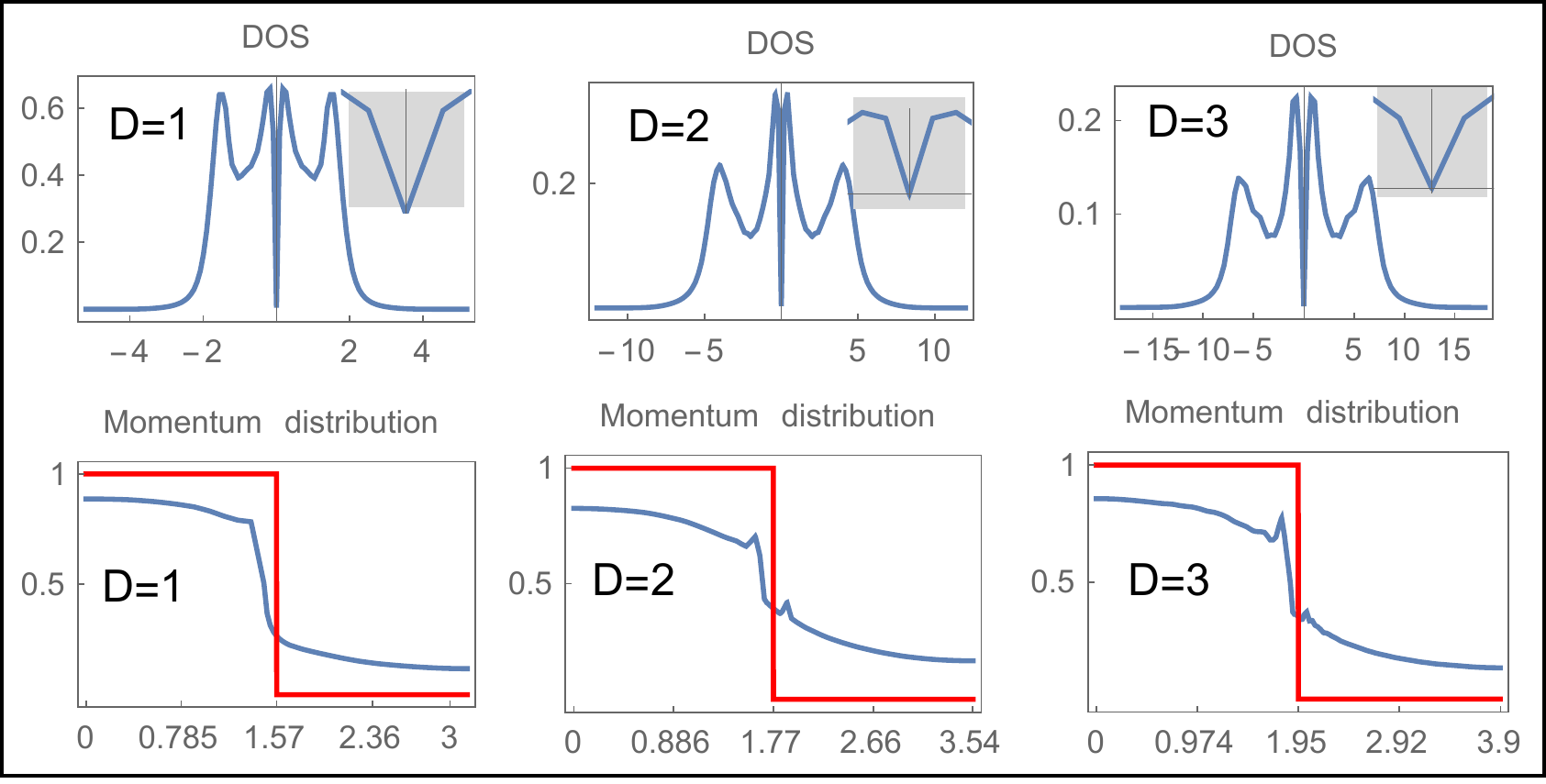}
\caption{Densities of electronic states and momentum distributions in the occupation number at $D=1, 2, 3$ for Coulomb interaction values $U\simeq U_{cD}$ that are near to the lattice distortion. The red step function is the momentum distributions of free electrons.}
\label{Pm:Dos_nk}
\end{center}
\end{figure}

Fig.4 represents DOS and momentum distributions  
in the occupation number at dimensions $D=1, 2, 3$\, when the system is near to CDW---lattice distortion, i.e. $U\simeq U_{cD}$. 
The upper series of pictures give DOS as a function of a frequency $\omega$ with corresponding insets for $\rho$ near  $\omega\sim 0$. We see that for all dimensions there is no finite gaps in DOS, but $\rho(0)\approx 0$; therefore the ground state of the system is semimetallic. The bottom series of pictures represent the momentum distributions $n(k)$ (\ref{DPM:nk}).  
The functions whithout any finite jumps at the Fermi momentums (\ref{PM:KF}) $k_F\approx 1.57 (D=1)$, $k_F\approx 1.77 (D=2)$, 
$k_F\approx 1.95 (D=3)$ do not exhibit the Fermi liquid behavior of the system, though with $D$ inreasing 
we observe some tendency toward it.

\section{\!\!\!\!\!\!. Conclusion}
 
In this paper we have studied the problem of the lattice dimension influence on the electronic properties of the half--filled paramagnetic HM at zero temperature. The model was formalized in terms of integral equations obtained previously 
\cite{{chaschin2011_2},{chaschin2012_3},{chaschin2011_1}}. We supposed for the hypercubic lattice of an arbitrary dimension $D$ all relevant functions depend only on the magnitude of momentum vectors 
(\ref{Spectrum_D}) and then choosed the free electronic spectrum as in the strong-coupling regime 
(\ref{D_freespect}). As a consequance all lattice sums were written as the $D-$space integrals 
(\ref{D_k}), (\ref{PM:Kd}) over the first BZ of corresponding radius $K_D$. Accordingly, after some transformations we got the HM model, where the dimension is the external parameter and not necessarily an integer one 
(\ref{Sigma}), (\ref{Pi_Q}).  

Firstly make a note that Coulomb interaction $U$ appeares in the model equations as a factor of the multiplication $U C_{D-1}$ (\ref{CD}), so this is the real constant parameter of the model. From Fig.1 we see that $C_{D}\ll1$ for D>4 and this indicates that $D=4$ is some number threshold for correlation effects --- they are perceptible only for a quite large U.

Fig.3 represents graphics of the number of double--occupied sites as a function of Coulomb interaction at different $D$. With $U$ growth $\langle n_{\uparrow}n_{\downarrow}\rangle$ is decreasing to zero at some value of $U=U_{cD}$ because of CDW.  A negative value of this parameter for $U>U_{cD}$ causes the CDW---instability that brings the lattice to a period doubling with respect to local charge values ("holes" and "twos"); that means a phase transition  with $\langle n_{\uparrow}n_{\downarrow}\rangle$ as the order parameter.

The graphics of electronic states and momentum distributions in the occupation number at the different space dimensions are shown in Fig.4. For Coulomb interaction close to $U_{cD}$ there is no finite jumps at the Fermi momentums $k_{FD}$ and finite gaps in DOS excepting one point at $\omega=0$, therefore the system does not exhibit the Fermi liquid behavior.

Futher stage of investigations in this direction require considering of a possibility of two sublattice structures, which as we suppose would appear at $U>U_{cD}$. 

\section*{Appendix}
\addcontentsline{toc}{chapter}{Appendix}
\setcounter{equation}{0}
\section*{General formulas for a spatially homogeneous paramagnetic solution}
\addcontentsline{toc}{section}{General formulas for a spatially homogeneous paramagnetic solution}

It had been shown in \cite{chaschin2011_2}
 % based on equations obtained within the generating functional%method [11--15] with the subsequent 
%Legendre transformation [9, 10] 
that in the configuration representation Hubbard model is determined by the following set of equations: 

\begin{equation}
\left\{
\begin{array}{l}
\displaystyle (G_0^{-1} \,N)(12)+\left[h(11)+U\, M(11)\right] 
M(12)=2\, \delta(12)-U\,X(12\,;11) 
\\
{}
\\
(G_0^{-1} \,M)(12)=-\left[h(11)+\frac{U}{2}\,M(11)\right] N(12)
\\
{}
\\
\displaystyle X(12;34)+\frac{U}{2}\,X(12;1^\prime 1^\prime)\, G_0(1^\prime 4)\,N(31^\prime)= -\,
\displaystyle G_0(14)\,N(32), 
\end{array}
\right.
\tag{A.1}
\label{Gen:eqn}   
\end{equation}
where $1\equiv (\tau, {\bf R})$, $\tau$ is the imaginary thermodynamic time and ${\bf R}$ is a lattice site; $G_0$ is Green's function in the Hartree--Fock approximation (HFA) --- free electronic propagator in the model;  $N=G_\uparrow+G_\downarrow$ and $M=G_\uparrow-G_\downarrow$ are propagators of the number of particles and magnetic moment respectively; h(11) is the local magnetic field. Henceforth we put the spontaneous local moment $M(11)=0$. The primed indexes are supposed to be summed.

The multiparticle correlator $X(12\,;34)$, which is responsible for
the electron--electron correlations in the system obeys a Bethe--Salpeter --- like equation
\begin{equation}
\displaystyle X(12\,;34)+\frac{U}{2}\,G_0(1 1^\prime)\,N(1^\prime 2)
X(1^\prime 1^\prime;34)\, =-\,\displaystyle G_0(14)\,N(32)\,. 
\tag{A.2}
\label{Gen:X}
\end{equation}

In the half--filled symmetrical case ($\displaystyle \frac{U}{2}-\mu=0$,\; $N(11)=\langle n\rangle-1=0$)\, for   
the inverse HFA propagator we have  

\begin{equation}
\begin{array}{l}
\displaystyle G_0^{-1}(12)=-\frac{d}{d\tau}\,\delta(12)-t(12)\,;\\
\displaystyle G_0^{-1}({\bf k})=i\,\omega_n-\varepsilon_{\bf k},\,\,\omega_n=(2 n+1)\pi T\,.
\end{array}
\tag{A.3}
\label{Gen:G_0}
\end{equation} 

The Fourier transform of the corresponding one--particle correlator $X(12\,;11)$ in the spatially homogeneous case reads 
\begin{equation}
\displaystyle X(12;11)=\sum\limits_k e^{i k(1-2)}\,X(k)\,, \quad
\displaystyle X(k)=-N_k\sum\limits_q \,G_0(q+k)Q_q\:,
\tag{A.4}
\label{Gen:X_f}
\end{equation}
where  
\begin{equation}
\displaystyle Q_q=\frac{1}{1+\frac{U}{2}
\displaystyle \sum\limits_{k} G_0(k+q)\,N_k }\:\,. 
\tag{A.5}
\label{Gen:Q}
\end{equation}
The notations k, q here represent the combined symbol (${\bf k}, i\omega_n$) and 
$\sum\limits_{k}\dots$ denotes the simultaneous summation
over the momentum vector ${\bf k}$ and imaginary discrete frequency $i\omega_n$.

After the  Fourier transformation of Eqs. (\ref{Gen:eqn}) and taking into consideration 
the expressions (\ref{Gen:X_f}, \ref{Gen:Q}) we get the next set of equations: 
\begin{equation}
\left\{
\begin{array}{l}
\displaystyle N_k=\frac{2}{G^{-1}_0(k)-\Sigma_k}\,,\\{}\\
\displaystyle \Sigma_{k} = \sum_{q}G_0(q+k)Q_q\,,\quad 
\displaystyle \Pi_q = \sum_k  G_0(k+q)N_k\,.
\end{array}
\right.
\tag{A.6}
\label{Ap:N}   
\end{equation}

The analytical continuation from the range of imaginary frequencies to the upper half-plane, i.e.  
$i\omega_n\to\omega+i0$, $i\Omega_\nu\to\Omega+i0$, and then subsequent standard spectral transformations including the Kramers--Kr\"{o}nig relation of the relevant functions 
make it possible to represent the spatially homogeneous half--filled paramagnetic HM in the form of the following sets of integral equations:

\begin{equation}
\left\{
\begin{array}{c}
\displaystyle \Im\Pi({\bf q},\Omega) = \frac{U}{2}\sum\limits_{{\bf k}}
\displaystyle \left[1-\tanh(\frac{\varepsilon_{{\bf k}}}{2 T})
\displaystyle \tanh(\frac{\varepsilon_{{\bf k}}-\Omega}{2 T})\right]
\Im N({\bf k}-{\bf q},\varepsilon_{{\bf k}}-\Omega)\,,
\\
{}
\\
\displaystyle \Re \Pi({\bf q},\Omega) = \frac{1}{\pi}\int_{-\infty}^\infty  
\displaystyle \frac{\tanh(\frac{\Omega^\prime}{2 T})\Im \Pi({\bf q},\Omega^\prime)}{\Omega^\prime - \Omega}\,
d\Omega^\prime\,, 
\\
{}
\\
\Im Q({\bf q};\Omega)=\displaystyle\frac
{2 \,\Im \Pi({\bf q},\Omega)}{\left[2+\Re\Pi({\bf q},\Omega)\right]^2+
\left[\Im \Pi({\bf q},\Omega)\tanh(\frac{\Omega}{2 T})\right]^2}\,; 
\end{array}
\right.
\tag{A.7}
\label{Ap:PI_Q}
\end{equation}

\begin{equation}
\left\{
\begin{array}{c}
\displaystyle \Im \Sigma({\bf k},\omega) = \frac{U}{2}\sum\limits_{{\bf q}}
\displaystyle \left[1-\tanh(\frac{\varepsilon_{{\bf q}}}{2 T})
\displaystyle \tanh(\frac{\varepsilon_{{\bf q}}-\omega}{2 T})\right]
\displaystyle\Im Q({\bf q}-{\bf k} ,\varepsilon_{{\bf q}}-\omega)\,,
\\
{}
\\
\displaystyle \Re \Sigma({\bf k},\omega) = \frac{1}{\pi}\int_{-\infty}^\infty  
\displaystyle \frac{\Im \Sigma({\bf k},\omega^\prime)}{\omega^\prime - \omega}\,
d\omega^\prime\,, 
\\
{}
\\
\Im N({\bf k},\omega)=\displaystyle\frac
{2\,\Im \Sigma({\bf k},\omega)}{\left[\omega-\varepsilon_{\bf k}-\Re\Sigma({\bf k},\omega)\right]^2+
\left[\Im \Sigma({\bf k},\omega)\right]^2}\,. 
\end{array}
\right.
\tag{A.8}
\label{Ap:Sigma_N}
\end{equation}

It can easily be shown by the direct substitution that Eqs. (\ref{Ap:PI_Q},\ref{Ap:Sigma_N}) are under the following symmetry conditions: 
 \begin{equation}
\left\{
\begin{array}{l}
\displaystyle \varepsilon({\bf k}+{\bf K}_D) = -\varepsilon({\bf k});\\ 
\displaystyle \Im\Sigma({\bf k}+{\bf K}_D,-\omega) = \Im\Sigma({\bf k},\omega), \quad 
\Re\Sigma({\bf k}+{\bf K}_D,-\omega) = -\Re\Sigma({\bf k},\omega); \\
\displaystyle \Im N({\bf k}+{\bf K}_D,-\omega) = \Im N({\bf k},\omega),\quad 
\Re N({\bf k}+{\bf K}_D,-\omega) = -\Re N{\bf k},\omega); \\
\displaystyle\Im\Pi({\bf q}+{\bf K}_D,-\Omega) = \Im\Pi({\bf q},\Omega),\quad  
\Re \Pi({\bf q}+{\bf K}_D,-\omega) = \Re \Pi({\bf q},\Omega)\,;\\
\displaystyle\Im Q({\bf q}+{\bf K}_D,-\Omega) = \Im Q({\bf q},\Omega),\quad 
\displaystyle\Re Q({\bf q}+{\bf K}_D,-\Omega) = \Re Q({\bf q},\Omega).\quad  
\end{array}
\right.
\tag{A.9}
\label{Ap:Symm}   
\end{equation}

Just to make (A.7, A.8)  structurally more identical we have put more convenient notation for 
$\Im\Pi({\bf q}, \Omega)=-U\,\coth(\frac{\Omega}{2 T})\Im\Pi({\bf q}, \Omega+i 0)$, as compared with one in 
\cite{chaschin2016_4}.

\end{document}